\documentclass[sigconf,nonacm]{acmart}

\usepackage{tabularx}
\usepackage{multirow}

\usepackage{fontawesome5}
\newcommand{\graphOnly}{\faTimes}

\usepackage{tikz}
\usetikzlibrary{positioning,arrows.meta,shapes.multipart}

\tikzset{
  pg node/.style = {
  rectangle split,
  rectangle split parts = 2,
  rectangle split part align = {center,left},
  draw = black,
  line width = 0.5pt,
  rounded corners = 1pt,
  inner xsep = 4pt,
  inner ysep = 3pt,
  minimum width = 22mm,
  font = \footnotesize,
  align = left
  }, 
  pg edge/.style = {
    ->,
    >= {Stealth[length = 5pt,width = 5pt]},
    line width = 0.6pt
  },  
  pg edge label/.style = {
    fill = white,
    inner xsep = 2pt,
    inner ysep = 1pt,
    font = \scriptsize,
    align = center 
  }, 
  pg missing node/.style = {
    pg node,
    dashed
  },
  pg missing edge/.style = {
    pg edge,
    dashed
  }
}

\newcommand{\mapping}[1]{%
  \par\noindent\hspace*{1em}%
  {\normalfont\scshape Mapping: }{\normalfont #1}%
}

\usepackage{circledsteps}
\newcommand{\myCircled}[4]{%
  \pgfkeys{/csteps/inner color=#1}%
  \pgfkeys{/csteps/outer color=#2}%
  \pgfkeys{/csteps/fill color=#3}%
  \Circled{#4}%
}
\definecolor{amber}{rgb}{1.0, 0.75, 0.0}
\newcommand{\cyYellow}[1]{\myCircled{black}{amber!70!black}{amber!25}{#1}}

\AtBeginMaketitle{%

\def\ISSN{xxxx-xxxx}
\def\ISBN{xxx-x-xxxxx-xxx-x}
\setcopyright{none}
\copyrightyear{\copyright\ 2027 Copyright held by the owner/author(s).
  Published on xxxxx under ISBN \ISBN, series ISSN \ISSN.
  Distribution of this paper is permitted under the terms of the
  Creative Commons license CC-by-nc-nd 4.0}
\acmYear{2027}
\acmDOI{}
\acmISBN{}

\acmConference[ACRONYM]{NAME}{DATE}{PLACE}

\settopmatter{printacmref=false, printccs=false, printfolios=false}

\newsavebox{\ximagebox}
\newlength{\ximageheight}
\newsavebox{\xglyphbox}
\newlength{\xglyphheight}
\newcommand{\xbox}[1]%
  {\savebox{\ximagebox}{#1}%
  \settoheight{\ximageheight}{\usebox{\ximagebox}}%
  \savebox{\xglyphbox}{\color{white}\char32}%
  \settoheight{\xglyphheight}{\usebox{\xglyphbox}}%
  \raisebox{\ximageheight}[0pt][0pt]{\raisebox{-\xglyphheight}[0pt][0pt]{%
    \makebox[0pt][l]{\usebox{\xglyphbox}}}}%
    \usebox{\ximagebox}%
    \raisebox{0pt}[0pt][0pt]{\makebox[0pt][r]{\usebox{\xglyphbox}}}}

\newsavebox{\LogoBox}
\sbox{\LogoBox}{}

\makeatletter
\fancypagestyle{firstpagestyle}{%
  \fancyhf{}%
  \fancyfoot{}%
  \fancyhead[L]{\ACM@linecountL}
  \fancyhead[R]{%
    \raisebox{15pt}[0pt][0pt]{\xbox{\usebox{\LogoBox}}}%
  \ACM@linecountR
  }%
}
\makeatother

\hypersetup{%
  pdfcopyright={CC-by-nc-nd},
  pdfcreator={LaTeX with acmart, hyperref, and modifications},
  pdfvolumenum={XX},
  pdfissuenum={X},
  pdfisbn={\ISBN},
  pdfissn={\ISSN}
}

}

\begin{document}

\title{An Error Taxonomy for Property Graph Data}

\author{Valerie Restat}
\email{valerie.restat@fernuni-hagen.de}
\orcid{0000-0002-5960-5886}
\affiliation{%
  \institution{FernUniversit\"{a}t in Hagen}
  \city{Hagen}
  \country{Germany}
}

\author{Dominique Hausler}
\orcid{0009-0004-2381-133X}
\email{dominique.hausler@ur.de}
\affiliation{%
  \institution{University of Regensburg}
  \city{Regensburg}
  \country{Germany}
}

\author{Désirée Bruns}
\orcid{0009-0008-6895-857X}
\email{desiree.bruns@studium.fernuni-hagen.de}
\affiliation{%
  \institution{FernUniversit\"{a}t in Hagen}
  \city{Hagen}
  \country{Germany}
}

\author{Uta Störl}
\orcid{0000-0003-2771-142X}
\email{uta.stoerl@fernuni-hagen.de}
\affiliation{%
  \institution{FernUniversit\"{a}t in Hagen}
  \city{Hagen}
  \country{Germany}
}

\renewcommand{\shortauthors}{Restat et al.}

\begin{abstract}
Graph data are increasingly used in applications where relationships and topology matter. Yet, compared to tabular data, the notions of data quality and error types for graphs are less established. Tabular error concepts often fail to capture structural properties and context inherent in graph data, hindering systematic quality assessment in graph-based systems.
We focus on property graphs (PGs) in graph databases. PGs are often managed without a prescriptive schema, enabling data insertion without explicit validation, while still exhibiting an implicit schema of recurring structural patterns. Irregularities in these patterns may indicate errors, but their interpretation depends on the surrounding graph context. 
To address these challenges, we provide a systematic view of errors in graph data by proposing a comprehensive graph-specific taxonomy of error types. To position these errors with respect to established work on tabular data quality, we derive a mapping between graph errors and corresponding error notions in tabular data. This mapping highlights which concepts transfer across data models and where structural properties of graphs introduce fundamentally different challenges requiring extensions. Overall, the taxonomy and mapping provide a shared vocabulary for describing, characterizing, comparing, and studying graph data errors, and they lay the groundwork for future work on error detection and repair in graph data.
\end{abstract}


\begin{teaserfigure}
 \includegraphics[width=\textwidth]{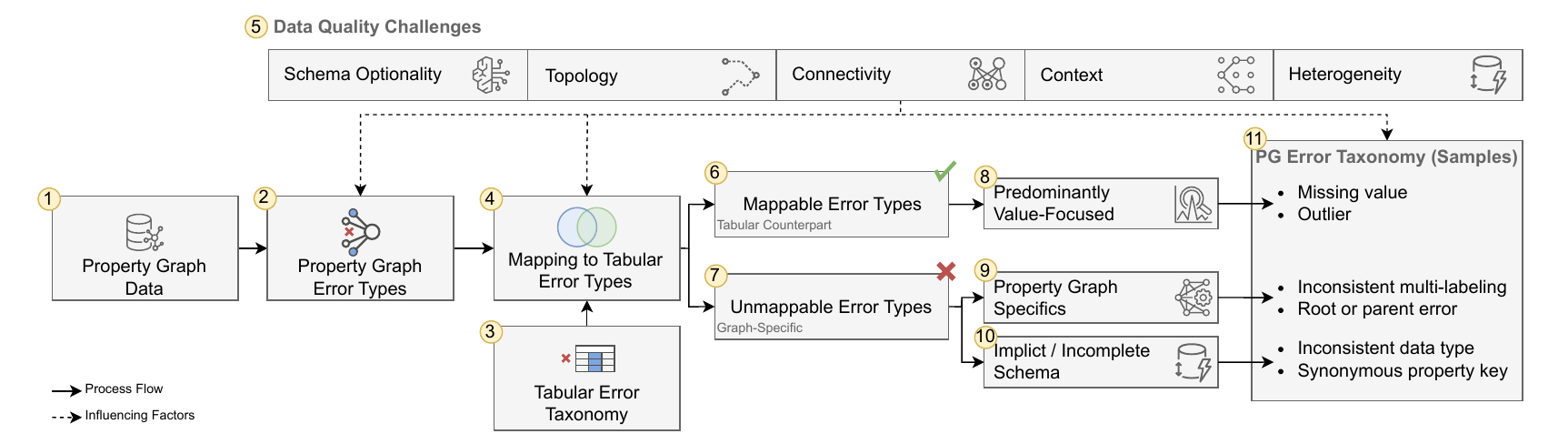}
  \caption{Approach toward a Property Graph Error Taxonomy under the overarching influence of data quality challenges. }
  \Description{A diagram illustrating the taxonomy of property graph errors and their relation to data quality.}
  \label{fig:teaser}
\end{teaserfigure}

\maketitle

\section{Introduction}
\label{sec:intro}
Graph data are widely used in domains where relationships between data points play a central role. These include social networks, recommender systems or fraud detection, as they naturally capture highly interconnected data. As graph-based systems become more pervasive~\cite{Sakr2021}, ensuring the quality of their data grows in importance. Errors in graph data can significantly alter the reliability of applications, leading to misinterpretations and wrong decisions~\cite{Wang2021}. 

We particularly focus on property graphs (PGs), which model entities as nodes and relationships as edges, with both carrying labels and properties. Therefore, errors in PGs involve not only properties and labels but also existence and connectivity of nodes and edges as well as larger structural patterns in the graph. 
This illustrates the need for a PG taxonomy, which is currently missing.

Even in the absence of an explicit schema, the underlying graph data still contain an implicit schema, following recurring structural patterns \cite{DBLP:conf/er/BonifatiFGHOV19}. Variations from these patterns serve as error indicators, but do not necessarily constitute errors themselves \cite{DBLP:conf/lwa/MollerHSAK23}. We use the terms explicit and implicit schema to distinguish externally specified schema information from inferred schema information, respectively. 

Systematic error detection requires an understanding of what constitutes an error in the first place. For tabular (relational) data, numerous works have organized established notions into error taxonomies and catalogs of error types and indicators (e.g.,~\cite{Bhadauria2026}). In contrast, these taxonomies do not cover graph-specific, topology-dependent errors in PGs. In this paper, we take a first step toward closing this gap by developing a technology-independent PG error taxonomy. To structure it, we build on the levels of occurrence proposed by Oliveira et al.~\cite{Oliveira2005} and distinguish property-, entity-, relation-, graph-, and pattern-level errors.

Figure~\ref{fig:teaser} illustrates our approach to deriving such a technology-independent PG error taxonomy. By analyzing PG data in Step~\cyYellow{1} corresponding error types are identified (Step~\cyYellow{2}). While most research focuses on data quality in structured, tabular data (Step~\cyYellow{3}), transferring existing error types to graph data remains challenging (Step~\cyYellow{4}). Graphs differ fundamentally due to their flexible or even schema-free nature and because their topology carries meaning. Several data quality challenges, such as connectivity or heterogeneity, directly influence the design of our taxonomy (Step~\cyYellow{5}). In graph data, two cases emerge: 1) a direct mapping of error types is possible (Step~\cyYellow{6}) which accounts for predominantly value-focused error types (Step~\cyYellow{8}) or 2) the graph structure itself becomes part of what may be erroneous, missing, inconsistent, or semantically misleading (Steps~\cyYellow{9} and~\cyYellow{10}). The latter represent error types without a tabular counterpart (Step~\cyYellow{7}), thus being unmappable. Due to the inherent connectivity and the semantic importance of an element's surroundings, these characteristics demand different approaches to assessing and improving data quality, since some errors cannot be assessed based on individual graph elements alone and require consideration of their structural context. Together, the identified error types form our PG error taxonomy, with a subset of examples shown in Step~\cyYellow{11}.

\paragraph{Contribution.} With this paper, we aim to make the following contributions: 
\begin{itemize}
    \item \textit{Technology-independent PG error taxonomy}: To the best of our knowledge, we present the first systematic, technology-independent error taxonomy for property graphs.
    \item \textit{Explicit bridge to tabular error taxonomies}: We are the first to systematically relate PG error types to established tabular error taxonomies. Unlike other studies, in which the various error types are always considered in isolation, we bridge the gap between the different data models. In doing so, we highlight the commonalities and identify where structural differences result in error types that exist only in either graph data or tabular data.
\end{itemize}

While we provide a comprehensive error taxonomy and a mapping to tabular error types, we deliberately limit our scope to error characterization and do not cover methods for error detection or repair in graph data. The systematic characterization on what constitutes an error and where errors manifest which we presented in this paper builds the conceptual foundation for developing and evaluating such methods. 
Detection and repair introduce additional challenges (e.g., algorithm design, domain-specific assumptions, and evaluation) that are beyond the scope of this work. We therefore leave the development of error detection and repair for future work.

\paragraph{Outline.} The remainder of the paper is structured as follows: First, we discuss relevant literature in Section~\ref{sec:related_work}, followed by the challenges emerging from working with graph data (Section~\ref{sec:dq-challenges}). Section~\ref{sec:errors} focuses on our universal error taxonomy for PGs, including a mapping to the aligning tabular counterparts. Afterwards, Section~\ref{sec:discussion} revisits key characteristics and challenges of graph data and outlines how they shape error detection and repair, before we conclude in Section~\ref{sec:conc}.

\section{Related Work}
\label{sec:related_work}
In this section we analyze existing approaches to error taxonomies, graph data quality as well as graph schema and constraints. 

Data quality has been a subject of research for over 30 years~\cite{Wang1993, Wang1995}. Various approaches have been developed to classify errors into taxonomies. In this context, in addition to the term \textit{data error}, terms such as \textit{data problems}, \textit{data issues}, or \textit{dirty data} are also frequently used. This paper uses the terms \textit{data error} and \textit{error taxonomy}.

\paragraph{Error Taxonomies.} A commonly used taxonomy is that of Rahm and Do~\cite{Rahm2000}. They use the term \textit{data quality problems} and classify these into single-source and multi-source problems, as well as schema-level and instance-level problems. Kim et al.~\cite{Kim2003} use the term \textit{dirty data} to refer to data errors. They divide these into three categories: \textit{missing data}, \textit{data that is not missing but incorrect}, and \textit{data that is neither missing nor incorrect but unusable}. These categories were then further divided into subcategories, which in turn encompass the specific types of data errors. Li et al.~\cite{Li2011} also use the term \textit{dirty data} and classifies errors into four rule categories: business entity rules, business attribute rules, data dependency rules,  and data validity rules. Oliveira et al.~\cite{Oliveira2005} -- like Rahm and Do~\cite{Rahm2000} -- use the term \textit{data quality problems}. Their taxonomy follows a \textit{bottom-up approach}, in which data errors are categorized into six levels. As with Rahm and Do~\cite{Rahm2000} and Kim et al.~\cite{Kim2003}, a distinction is made between single-source and multi-source errors. Single-source errors can be divided into \textit{Single Relation} and \textit{Relationships among Multiple Relations}. The group \textit{Single Relation} includes the four lowest levels \textit{An Attribute Value of a Single Tuple}, \textit{The Values of a Single Attribute}, \textit{The Attribute Values of a Single Tuple}, and \textit{The Attribute Values of Several Tuples}. In~\cite{GouDa}, we extended this taxonomy to incorporate additional important errors, based on~\cite{Rahm2000, Kim2003, Li2011}. Quindroit et al.~\cite{Quindroit2023} survey a broad range of error taxonomies. Among the taxonomies with hierarchies, the most comprehensive are those by Kim et al. (33 items)~\cite{Kim2003}, Oliveira et al. (30 items)~\cite{Oliveira2005}, and Li et al. (38 items)~\cite{Li2011}. In our taxonomy~\cite{GouDa}, we combine these three (and that of Rahm and Do~\cite{Rahm2000}) to cover a broad spectrum of error types across various sources and levels.
One particular new taxonomy is that proposed by Bhaduria et al.~\cite{Bhadauria2026}, who use the term \textit{data error}. Their work proposes a catalog of 35 distinct error types for tabular data. The catalog distinguishes between \textit{data error types} and \textit{error indicators}. Similar to previous work, the authors organize data errors according to table structures, such as \textit{Value} and \textit{Tuple}. In addition, the error types are classified as \textit{missing}, \textit{incorrect}, or \textit{redundant}.

\textit{While these taxonomies provide a useful starting point, they were developed for tabular data, which differs structurally from graph data. Specifically, they do not account for topology and structural relationships. Thus, existing error taxonomies need to be systematically reevaluated to determine which concepts transfer to graph data and which graph-specific errors need to be included.}

Senaratne et al.~\cite{Senaratne2025} propose a taxonomy of anomaly types in knowledge graphs. They distinguish between anomalies in entity-based triples, anomalies in literal-based triples, and anomalies involving both types of triples.

\textit{Nevertheless, the taxonomy is limited to RDF-based knowledge graphs, covers few anomaly types and does not provide detailed description nor a mapping to tabular data errors.}

\paragraph{Graph Data Quality.} A growing body of survey literature has explored different aspects of graph data quality. Zaveri et al.~\cite{Zaveri2016} focus on quality assessment and metrics for RDF-based linked data. More recently, Cortés et al.~\cite{DBLP:conf/wop/CortesEEN25} investigate to what extent SHACL constraints can operationalize data quality metrics for knowledge graphs. Paulheim~\cite{Paulheim2016} surveys knowledge graph refinement methods, especially for adding missing knowledge, and identifying or removing errors. Wang et al.~\cite{Wang2021} as well as Xue and Zou~\cite{Xue2023} broaden this perspective towards knowledge graph quality control and quality management across the lifecycle of construction, assessment, problem discovery, and improvement. 

\textit{These surveys mainly address quality dimensions and metrics, methods of graph refinement, and strategies for managing graph data quality in knowledge graphs. Their focus is therefore on assessing and improving graph data quality, rather than systematically characterizing errors across the structural levels of a graph.
}

\paragraph{Graph Schema and Constraints.}
PGs offer direct data insertion and thus are associated with flexible schema management. Bonifati et al.~\cite{DBLP:conf/er/BonifatiFGHOV19} distinguish between descriptive and prescriptive schemas and analyzes schema validation alongside evolution for PGs. This built the foundation to formalize their schema within PG-Schema~\cite{Angles2023}, which introduced graph types while highlighting the still few schema management options. Complementary work on PG-Keys~\cite{Angles2021} extends PG-Schema by constraints to support data integration tasks. Building on constraint-based representations, Spinrath et al.~\cite{DBLP:journals/pvldb/SpinrathBE26a} analyze the repair of PGs under PG-constraints by modifying elements and topology to resolve constraint violations. Beyond schema and key constraints, several types of graph dependencies have been proposed to express integrity conditions over graph structures, including graph entity dependencies~\cite{DBLP:journals/is/LiuKZZBWGHF24} and graph generating dependencies \cite{DBLP:conf/cikm/ShimomuraY024}. An overview of graph functional dependencies, additionally aligning several of those them with PG-Schema, is given by Manouvrier and Belhajjame~\cite{DBLP:journals/is/ManouvrierB26}.

Since the introduction of the ISO/PGQ \cite{ISO_SQL_PGQ} and GQL \cite{ISO_GQL} standard, PG databases such as Neo4j consistently evolve the landscape of query languages to align with the standard, for instance, by integrating the \texttt{GRAPH TYPE} statements \cite{neo4j_graph_types}, first proposed in~\cite{Angles2023}. Related profiling approaches derive structural information directly from the graph data, ranging from RDF profiling and schema discovery~\cite{AbedjanGJN14} to profiling specifically tailored to PGs in the absence an explicit schema~\cite{DBLP:journals/jdiq/MaioloEM20}.

The schema-flexible nature of graph databases can result in irregularities, especially if no predefined schema is used or in collaborative environments. Hausler and Klettke~\cite{DBLP:conf/edbt/HauslerK26} investigate the detection and comparison of variation within graph data profiles. Rather than declaring structural irregularities as errors, heterogeneity is forwarded to support user-informed cleaning decisions, since domain knowledge is required to separation intentional from unintentional variants.

Additionally, work on consistency through normalization is available. Skavantzos and Link introduce normal forms for PGs based on graph-tailed functional dependencies to reduce redundancy and sources of inconsistencies \cite{DBLP:journals/vldb/SkavantzosL25}. More recent approaches extend this idea by proposing a graph-native approach \cite{DBLP:conf/edbt/SchrottJH27}.

\textit{Overall, prior work provides mechanisms to describe, infer, constrain, and normalize graph structures to identify specific constraint violations. Nevertheless, these are restricted to particular structural properties or inconsistencies rather than a complete error specification across different structural levels of PGs.}

\section{Data Quality Challenges for Graph Data}
\label{sec:dq-challenges}

The following section discusses the data quality challenges specific to PGs arising from their highly interconnected structure and other graph-specific characteristics. This includes the lack of a predefined schema (Section~\ref{subsec:schema-optionality}), the definition of missingness (Section~\ref{subsec:missingness}), topological considerations including the impact on connectivity (Section~\ref{subsec:topology-connectivity}), the consideration of context (Section~\ref{subsec:context}) alongside heterogeneity (Section~\ref{subsec:heterogeneity})). 

\subsection{Schema Optionality}
\label{subsec:schema-optionality}
Unlike relational databases, many graph data settings do not require a predefined schema to be declared in advance, thus, allowing direct data insertion. Therefore, schema information may remain implicit, be only partially formalized, or be more explicit through schemas and constraints~\cite{Candel2022, Angles2023}. The degree of schema explicitness in graph data can vary considerably. This flexibility is often considered a strength in evolving graph settings, since such settings support dynamic, agile environments~\cite{Angles2023}. Graphs can be developed iteratively, with new labels, properties, and relationships being introduced or modified as requirements evolve~\cite{Shah2020}.

However, schema optionality comes with specific challenges for data quality. The same set of facts may have synonymous or inconsistent naming, varying data types, or inconsistent relationships. Comparable problems are well known from schema integration, where naming and structural conflicts arise from different representations of the same object across sources~\cite{Rahm2000}. In graph settings, the variation may also concern more fundamental choices. Information can be represented, for example, as a node or a property, as a directed edge or via an intermediate node. A particularly important challenge under an implicit schema and heterogeneous representation concerns identity management. If identifiers, typing, or key properties are not used consistently, it cannot be unambiguously resolved whether several nodes represent the same entity or whether one node represents several entities. Moreover, in graph settings, information about a real-world entity may not be confined to a single node, but may also be distributed across adjacent edges and nodes. This is also reflected in the work on graph keys and identity constraints, where identifying conditions may need to combine property values with structural context~\cite{Fan2017, Angles2021}.

Overall, schema optionality and weakly governed structure complicate the formulation of structural expectations in graph data and thereby make consistency a central quality concern. This contributes to several error types in graph data.

\subsection{Missingness Semantics}
\label{subsec:missingness}
The meaning of missingness depends on the underlying world assumption. Under a \textit{Closed World Assumption (CWA)}, missing statements are interpreted as false relationships~\cite{Nickel2016}. Under an \textit{Open World Assumption (OWA)}, on the other hand, missing statements are treated as unknown rather than false~\cite{Galarraga2013}. Missingness in graph data therefore cannot be interpreted independently of the assumptions under which absence is evaluated. Whether a missing node or edge constitutes a genuine quality problem depends on the completeness assumptions in the respective graph setting.

\subsection{Topology \& Connectivity}
\label{subsec:topology-connectivity}
Graphs convey meaning not only through properties of single nodes or edges, but also through their topology. Seemingly minor alterations such as removing an edge or changing its direction can have non-local effects on connectivity, shortest paths, centrality, or community structures. This structural sensitivity is particularly consequential in graphs with a heavy-tailed degree distribution, where a small number of nodes have a large number of connecting edges. Such nodes are called hubs. Graphs with a small percentage of hubs can be highly vulnerable to changes affecting these structurally important nodes~\cite{Albert2000}. For example, removing just a tiny fraction of the nodes or edges in crime networks had a drastic impact on the structural similarity between the original graph and the pruned graph~\cite{Ficara2021}. 
Topological effects also change the character of familiar data errors. Duplicate or conflated structures are not merely analogous to tuples in a table. In a graph, they may introduce redundant neighborhood structures or establish connectivity between entities that should not exist caused by unintended entity merging.
Overall, topology means that even errors affecting only a single 
graph element may have wider consequences for the surrounding graph structure. Errors in graph data therefore cannot always be adequately understood at the level of isolated graph elements alone. Some become visible only through their effects on relations, neighborhoods, paths, or larger graph structures. This motivates the need for a graph-centric account of data quality.

\subsection{Context} 
\label{subsec:context}
While topology refers to the structural organization of a graph, more precisely how nodes and edges are connected and which structures result from these connections, context refers to the neighborhood of a considered graph element together with the label and property information needed to determine whether an element or structure is erroneous. 
Consequently, context aids in identifying irregularities by considering structural as well as label and property information. This information provides evidence as to whether observed heterogeneity constitutes an error indicator or whether it reflects an actual error. 
Such contextual information is highly relevant, since graph elements derive their semantics from their surroundings and cannot always be considered in isolation.
This becomes particularly important in the context of pattern-level errors, whose identification depends on comparing observed structures with recurring patterns in the surrounding graph. Since context differs in its scope, meaning the extent of the surrounding graph considered around the referenced element, it can range from the direct neighborhood (path length of 1) to larger paths or subgraphs. Additionally, domain context captures semantics that depend on either domain or external knowledge. For instance, values or relations can be consistent within the graph data but erroneous under real-world assumptions.

\subsection{Heterogeneity}
\label{subsec:heterogeneity}
The implicit schema captures information on recurring structural patterns that provides a basis for identifying irregularities. These irregularities can manifest across multiple levels, namely varying labels, properties or structural pattern, among otherwise homogeneous graph elements. Nevertheless, heterogeneity is not necessarily erroneous, as it may be intentionally created, but serves as an error indicator~\cite{DBLP:conf/edbt/HauslerK26}. External context thus provides additional evidence to distinguish intentional heterogeneity from unintentional variations that may indicate errors. 

\subsection{Considerations for the Taxonomy Design}
These graph-specific challenges influence how errors are characterized in PGs. In particular, they outline the required considerations when mapping graph errors to errors from existing taxonomies on tabular data. Additionally, it illustrates the necessity to account for structural errors beyond individual properties, distinguish irregularities from actual errors and consider the context in which a graph element or pattern occurs. These challenges therefore guide the organization and interpretation of error types and their mapping to tabular counterparts, presented in the following section. 

\section{Error Taxonomy}
\label{sec:errors}
To close the gap of a missing error taxonomy for graph data, outlined by Bhadauria et al.~\cite{Bhadauria2026}, we present a comprehensive error taxonomy for PGs across five levels. This is accomplished by conducting a mapping between our previous taxonomy for tabular data~\cite{GouDa} and its graph-specific counterparts. This allows us to identify how graph errors differ from tabular errors and supports the systematic identification of additional graph-specific error types. To ensure general applicability, the error taxonomy in this paper describes error types for PGs independently of a specific graph database system.
In this paper, we use \textit{tabular data} in a broad sense, covering both simple tables (e.g., CSV files) and relational, normalized as well as non-normalized data. We point out relevant differences where necessary. 

Our mapping builds on the tabular error taxonomy shown in Table~\ref{tab:tabular-taxonomy}. This tabular notion illustrates our error taxonomy~\cite{GouDa} which is based on the work of Oliveira et al.~\cite{Oliveira2005} and classifies errors according to their structural level. Moreover, the levels are partitioned based on their granularity with increasing complexity. Atomic error types are categorized at the \textit{An Attribute Value of a Single Tuple} level, since it is restricted to a single value in merely one tuple. \textit{The Values of a Single Attribute} refers to issues arising when analyzing a set of property values of one property. 
Errors affecting a set of values across multiple attributes are located at the level of \textit{The Attribute Values of a Single Tuple}. The last level called \textit{The Attribute Values of Several Tuples} refers to errors within a single relation and thus represents the highest level of complexity within a single instance. For the mapping, presented in this paper, we additionally include the level \textit{Relationships among Multiple Relations}, which was not listed in our previous taxonomy~\cite{GouDa}.

\begin{table}[!ht]
  \caption{Error taxonomy on tabular data by Oliveira~\cite{Oliveira2005}, expanded by Restat et al.~\cite{GouDa}}
  \label{tab:tabular-taxonomy}
  \begin{tabularx}{\columnwidth}{>{\bfseries}p{2cm} X}
    \toprule
    \textbf{Level} & \textbf{Error Type}\\
    \midrule
    \multirow{9}*{\shortstack[l]{An Attribute\\ Value of a\\ Single Tuple}} & V1 Missing value\\ 
     & V2 Syntax violation\\ 
     & V3 Interval violation\\
     & V4 Set violation\\ 
     & V5 Misspelled error\\ 
     & V6 Inadequate value to the attribute context\\
     & V7 Value items beyond the attribute context\\ 
     & V8 Meaningless value\\ 
     & V9 Erroneous entry\\
     & V10 Outdated value\\
    \midrule
    \multirow{4}*{\shortstack[l]{The Values\\ of a Single\\ Attribute}} & A1 Uniqueness value violation\\
     & A2 Synonyms existence\\
     & A3 Outlier\\ 
     & A4 Missing attribute\\
    \midrule
    \multirow{3}*{\shortstack[l]{The Attribute\\ Values of a\\ Single Tuple}} & T1 Semi-empty tuple\\
     & T2 Inconsistency among attribute values\\ 
     & T3 Irrelevant observation\\
    \midrule
    \multirow{4}*{\shortstack[l]{The Attribute\\ Values of\\ Several Tuples}} & S1 Redundancy about an entity\\
     & S2 Inconsistency about an entity\\
     & S3 Bias\\ 
     & S4 Noise\\
     \midrule
     \multirow{6}*{\shortstack[l]{Relationship\\ among\\ multiple\\ Relations}} & M1 Referential integrity violation\\
     & M2 Outdated reference \\ 
     & M3 Syntax inconsistency\\ 
     & M4 Inconsistency among related attribute values\\ 
     & M5 Circularity among tuples in a self-relationship \\ 
     & M6 Domain constraint violation\\
    \bottomrule
  \end{tabularx}
\end{table}

Oliveira et al.~\cite{Oliveira2005} also include a \textit{Multiple Data Sources} level which describes error types emerging if multiple sources are either integrated or used for cross-source analysis. Even though multi-source error types are out of the scope of this paper, the schema optionality of graph data can cause analogous issues within a single source or instance. Thus, our taxonomy already captures corresponding error types at different structural levels rather than introducing an additional level. We therefore focus on heterogeneity within a single PG instance, where schema optionality besides other data quality challenges (see Section~\ref{sec:dq-challenges}) already introduce substantial complexity, while a systematic analysis of multi-source errors is left for future work.

In line with this taxonomy, we have also divided the error types for graph data into different levels. Graphs, however, represent information not only through values, but through entities, relations, and larger subgraphs. Accordingly, our graph taxonomy distinguishes between the following levels: \textit{Property} (Section~\ref{subsec:property-level}), \textit{Entity} (Section~\ref{subsec:entity-level}), \textit{Relation} (Section~\ref{subsec:relation-level}), \textit{Pattern} (Section~\ref{subsec:pattern-level}), and \textit{Graph} (Section~\ref{subsec:graph-level}). The individual levels and their associated error types are described in the following sections.

These levels are not mutually exclusive in their consequences. An entity-level error may distort several relations, and a relation-level error may induce a pattern-level violation. Entries are nevertheless assigned to the level at which the violated condition can be defined most precisely. For the mapping, an error type is linked to a tabular counterpart when both capture an analogous data quality issue at the aligning structural levels.

For each level, a table depicts all associated error types. Within these tables we show the mapping to their tabular counterparts, whereby the left column lists the PG error types and the right one depicts the corresponding error type from Table~\ref{tab:tabular-taxonomy}. Not all graph error types can be mapped to a corresponding error type for tabular data. Whenever we could not identify a direct tabular counterpart in Oliveira et al., we first assessed whether one of the existing tabular error types could reasonably be transferred before classifying the issue as graph-specific.  The following sections, organized according to the graph-specific levels, explain the taxonomy and the mapping in greater detail and discuss which error types transfer across the tabular and the graph data models and which are specific to PGs.

\subsection{Property-level}
\label{subsec:property-level}
\begin{figure}[htbp]
\centering

\begin{tikzpicture}

\node[pg node] (n1) {
  \textbf{$n_1$: Track}
  \nodepart{second}
  title = 'Hit the Road Jack'\\
  duration = -2:00 (P3)
};

\node[pg node, right=15mm of n1] (n2) {
  \textbf{$n_2$: Track}
  \nodepart{second}
  title = 'Come on Back, Jack'\\
  length = 2:00 (P12)
};

\node[pg node, below=12mm of n1] (n3) {
  \textbf{$n_3$: Artist\,\textperiodcentered\,Person}
  \nodepart{second}
  name = 'Rae Charles' (P5)\\
  born = 23/09/1930 (P2)\\
  $\emptyset$ (P10)
};

\node[pg node, below=12mm of n2] (n4) {
  \textbf{$n_4$: Artist\,\textperiodcentered\,Person}
  \nodepart{second}
  name = 'Nina Simone'\\
  born = 1933/02/21\\
  died = 'April 21, 2003' (P17)
};

\draw[pg edge]
  (n2) to
  node[pg edge label, above, xshift=1mm] {
    \textbf{$e_3$: samples}
  }
  (n1);

\draw[pg edge]
  (n3) to
  node[pg edge label] {
    \textbf{$e_1$: performs}\\
    roll = 'lead vocals' (P11)
  }
  (n1);

\draw[pg edge]
  (n4) to
  node[pg edge label] {
    \textbf{$e_2$: performs}\\
    role = 'lead singing' (P14)
  }
  (n2);

\end{tikzpicture}

\caption{A subset of property-level error types.}
\Description{A figure showing a subset of property-level error types.}
\label{fig:property-level-graph}

\end{figure}
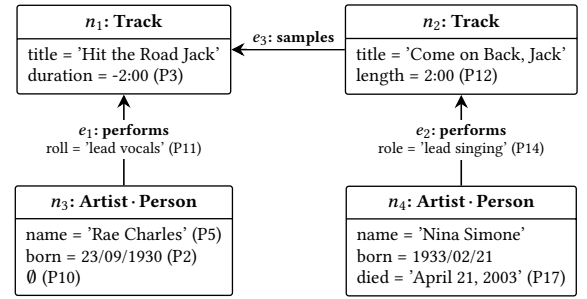

The first graph-specific level refers to properties.
Given that properties are represented as key-value pairs stored within each data point, namely node or edge, this category contains atomic error types. Consequently, error types classified at the \textit{Property-level} affect the key or value of a single property. In contrast, tabular data only specifies the attribute once for all values.
Figure~\ref{fig:property-level-graph} illustrates a subset of these errors using a sample graph from the music domain, while Table~\ref{tab:taxonomy-mapping-prop-level} contains all PG error types classified at the property-level including their tabular counterparts from Table~\ref{tab:tabular-taxonomy}. Error types without a corresponding tabular mapping are marked with \graphOnly. These error types are graph-specific.

\begin{table}[!ht]
  \caption{PG Error taxonomy for property-level error types. Entries marked with \graphOnly{} indicate graph-specific error types without a tabular counterpart.}
  \label{tab:taxonomy-mapping-prop-level}

  \small
  \setlength{\tabcolsep}{3pt}
  \renewcommand{\arraystretch}{1.15}

  \begin{tabularx}{\columnwidth}{%
    >{\raggedright\arraybackslash}X
    >{\raggedright\arraybackslash}X
  }
    \toprule
    \textbf{Graph Error Type} & \textbf{Tabular Error Type} \\
    \midrule
    P1 Missing value & V1 Missing value\\
    P2 Syntax violation & V2 Syntax violation\\
    P3 Interval violation & V3 Interval violation\\
    P4 Set violation & V4 Set violation\\
    P5 Misspelled error & V5 Misspelled error\\
    P6 Inadequate value to the property context & V6 Inadequate value to the attribute context\\
    P7 Value items beyond the property context & V7 Value items beyond the attribute context\\
    P8 Meaningless value & V8 Meaningless value \\
    P9 Erroneous entry & V9 Erroneous entry \\
    P10 Missing property key & \graphOnly \\
    P11 Misspelled property key & \graphOnly \\
    P12 Synonymous property key & \graphOnly \\
    P13 Missing property & A4 Missing attribute \\
    P14 Synonymous property value & A2 Synonyms existence \\
    P15 Uniqueness value violation & A1 Uniqueness value violation \\
    P16 Outlier & A3 Outlier \\
    P17 Inconsistency among data types & \graphOnly \\
    P18 Outdated value & V10 Outdated value\\
    P19 Outdated property key & \graphOnly \\
    \bottomrule
  \end{tabularx}
\end{table}

\paragraph{P1: Missing value} A property is missing the corresponding value for a key. We distinguish three manifestations of this error: First, a value defined as mandatory can be absent. Second, a value can be missing without being defined as mandatory, as indicated by an implicit schema. Third, a placeholder value such as \texttt{NULL} or an empty string can be used.  
\mapping{Same as the tabular error type \textit{Missing value (V1)}.}

\paragraph{P2: Syntax violation} The value of a property does not conform to the expected syntax. In Figure~\ref{fig:property-level-graph}, for instance, the value of the property key \texttt{born} in $n_3$ is formatted \texttt{dd/mm/yyyy}, whereas the corresponding value in $n_4$ follows \texttt{yyyy/mm/dd}. Such syntactic irregularities arise in schema-less PGs, where deviations from expected or implicit constraints are not violated or reported as errors.
\mapping{Same as the tabular error type \textit{Syntax violation (V2)}.}

\paragraph{P3: Interval violation} The value of a property is outside the allowed range. In Figure~\ref{fig:property-level-graph} this error type is represented by the negative value for the \texttt{duration} of a \texttt{Track} in $n_1$, which contradicts real-world knowledge. 
\mapping{Same as the tabular error type \textit{Interval violation (V3)}.}

\paragraph{P4: Set violation} In an enumeration, the value of a property conflicts with the set of valid values. If the property \texttt{releaseType} of a \texttt{Release} is restricted to the values \texttt{album, single, EP} or \texttt{audiobook} but instead stores the name of an actor. 
\mapping{Same as the tabular error type \textit{Set violation (V4)}.}

\paragraph{P5: Misspelled error} The value of a property is misspelled. Such syntactic variations represent typographical errors, e.g., \texttt{Rae Charles} instead of \texttt{Ray Charles}. In contrast to syntax errors, misspelled errors are limited to textual values.
\mapping{Same as the tabular error type \textit{Misspelled error (V5)}.}

\paragraph{P6: Inadequate value to the property context} The value of a property does not fit within its own context, but rather within that of another property. This error type solely consists in textual property values. Specifically, if both \texttt{title} and \texttt{category} properties are of type \texttt{String}, whereby \texttt{Back to Black} is stored as \texttt{category}, even though it should be stored as \texttt{title}. As indicated by the name their identification requires external knowledge. 
\mapping{Same as the tabular error type \textit{Inadequate value to the attribute context (V6)}.}

\paragraph{P7: Value items beyond the property context} Multiple values that are subsumed within single property value, even though parts of these values extend beyond the context of the property. Such substrings of textual values carry semantics associated with other, additional properties. For instance, the property \texttt{first\_name} stores the value \texttt{Peter Smith}, thus indicating a non-atomic value that subsequently shows normalization potential.
\mapping{Same as the tabular error type \textit{Value items beyond the attribute context (V7)}.}

\paragraph{P8: Meaningless value} The value of a property is out of context and cannot be assigned to any other property context.
\mapping{Same as the tabular error type \textit{Meaningless value (V8).}
}

\paragraph{P9: Erroneous entry} The value of a property that, while formally correct, does not correspond to the true value for the entity or relation. If \texttt{2027/12/31} is stored as the date for Christmas. 
\mapping{Same as the tabular error type  \textit{Erroneous entry (V9)}.}

\paragraph{P10: Missing property key} A property's key (and thus its associated value) is missing from an entity or a relation like \texttt{died} for $n_3$.
However, the property generally exists in other entities.
\mapping{Graph-specific: 
The property key is the counterpart of a column name. However, due to the explicitly defined schema
it cannot be missing from individual rows.  
}

\paragraph{P11: Misspelled property key} The key of a property is misspelled, such as \texttt{roll} instead of \texttt{role} of $e_1$ in Figure~\ref{fig:property-level-graph}.
\mapping{Graph-specific: This error illustrates unintentionally created heterogeneity which results from modeling flexibility and limited support for schema management.}

\paragraph{P12: Synonymous property key} A property's key contains a synonym. $n_1$, for example, uses \texttt{duration}, while $n_2$ uses \texttt{length} to store semantically identical values. 
\mapping{
Graph-specific: Schema-optionality can cause semantic variations, allowing unintended heterogeneity to remain undetected due to limited schema management.}

\paragraph{P13: Missing property} A property is systematically missing from the entire graph.
\mapping{Same as the tabular error type \textit{Missing attribute (A4)}.}

\paragraph{P14: Synonymous property value} A property's value contains a synonym. The \texttt{role} property of a \texttt{performs} relation is declared as \texttt{lead vocals} or \texttt{lead singing}, although both denote the same performer role.
\mapping{Same as the tabular error type \textit{Synonyms existence (A2)}.}

\paragraph{P15: Uniqueness value violation} Different entities or relations have the same value for the same property, even though the values of that property are supposed to be unique.
\mapping{Same as the tabular error type \textit{Uniqueness value violation (A1)}. However, it can only be directly identified as a violation under an explicit uniqueness constraint.}

\paragraph{P16: Outlier} An observation in a series of measurements that does not meet expectations. It is often necessary to refer to statistical distributions.
\mapping{Same as the tabular error type \textit{Outlier (A3)}.}

\paragraph{P17: Inconsistency among data types} More than one data type is used to store the values of the same property. Due to our taxonomy being technology-independent, the complexity arising depending on the allowed data types directly relates to JSON-specific considerations~\cite{DBLP:journals/tvcg/XiongXFWWW24}.
In $n_4$ of Figure~\ref{fig:property-level-graph} the \texttt{died} date is of type string.
\mapping{Graph-specific (w.r.t. Oliveira et al.~\cite{Oliveira2005}): Not covered in their relational taxonomy, since data types are explicitly defined in the schema. Note that in schema-less tabular settings (e.g., CSV files without an enforced schema), the same error can occur.
}

\paragraph{P18: Outdated value} A property value was not updated accordingly, thus storing an older data version. More precisely, this can occur when a working title was used for a music album and later changed before its release. Such an error requires external context to be identified.  
\mapping{Same as the tabular error type \textit{Outdated value (V10)}.}

\paragraph{P19: Outdated property key} A property key was not updated accordingly, thus storing an older version. 
\mapping{Graph-specific: Can result from lazy data migration in schema-optional systems, where outdated and updated property keys coexist within the same graph, causing temporal irregularities.}

\subsection{Entity-level}
\label{subsec:entity-level}
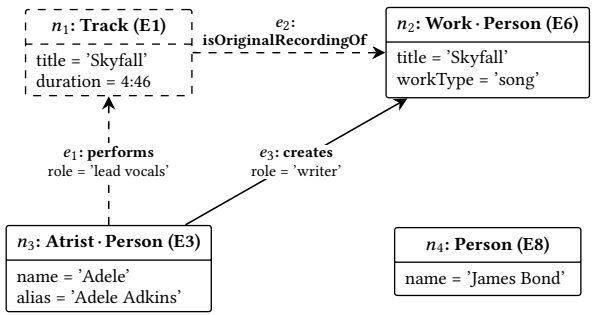
\begin{figure}[h!]
\centering

\begin{tikzpicture}

\node[pg missing node] (n1) {
  \textbf{$n_1$: Track (E1)}
  \nodepart{second}
  title = 'Skyfall'\\
  duration = 4:46
};

\node[pg node, right=25.5mm of n1] (n2) {
  \textbf{$n_2$: Work\,\textperiodcentered\,Person (E6)}
  \nodepart{second}
  title = 'Skyfall'\\
  workType = 'song'
};

\node[pg node, below=17mm of n1] (n3) {
  \textbf{$n_3$: Atrist\,\textperiodcentered\,Person (E3)}
  \nodepart{second}
  name = 'Adele'\\
  alias = 'Adele Adkins'
};

\node[pg node, below=17mm of n2] (n4) {
  \textbf{$n_4$: Person (E8)}
  \nodepart{second}
  name = 'James Bond'
};

\draw[pg missing edge]
  (n1) to
  node[pg edge label, above, xshift=-0.5mm] {
    \textbf{$e_2$:}\\\textbf{isOriginalRecordingOf}
  }
  (n2);

\draw[pg missing edge]
  (n3) to
  node[pg edge label] {
    \textbf{$e_1$: performs}\\
    role = 'lead vocals'
  }
  (n1);

\draw[pg edge]
  (n3) to
  node[pg edge label] {
    \textbf{$e_3$: creates}\\
    role = 'writer'
  }
  (n2);

\end{tikzpicture}

\caption{Examples of entity-level error types.}
\Description{A figure showing examples of entity-level error types.}
\label{fig:entity-level-graph}

\end{figure}

Given that nodes store properties represented as key-value pairs for each data point, the corresponding level, i.e., the \textit{Entity-level}, captures errors that arise from the consideration of a node or an edge as a whole. The term label is used interchangeably for node labels and edge types throughout our notion. Entity-level errors are those whose violated condition can be stated with respect to the representation of an entity, in particular its existence, identity, or classification. Note that both nodes and edges may carry labels. To avoid redundant descriptions, we subsume all label-related error types under the entity-level. However, these error types apply equally to relation labels. Analogously to the property-level, the sample graph in Figure~\ref{fig:entity-level-graph} uses the music domain to illustrate a subset of entity-level error types. Dashed lines thereby depict missingness. Table~\ref{tab:taxonomy-mapping-entity-level} shows all corresponding error types together with the mapping to their tabular version from Table~\ref{tab:tabular-taxonomy}, if applicable. 

\begin{table}[h!]
  \caption{Error taxonomy for entity-level errors in graph data. Entries marked with \graphOnly{} indicate graph-specific error types without a tabular counterpart.}
  \label{tab:taxonomy-mapping-entity-level}

  \small
  \setlength{\tabcolsep}{3pt}
  \renewcommand{\arraystretch}{1.15}

  \begin{tabularx}{\columnwidth}{%
    >{\raggedright\arraybackslash}X
    >{\raggedright\arraybackslash}X
  }
    \toprule
    \textbf{Graph Error Type} & \textbf{Tabular Error Type} \\
    \midrule
    E1 Missing (connected) entity & \textit{Missing tuple} \\
    E2 Missing label & \graphOnly \\
    E3 Misspelled label & \graphOnly \\
    E4 Synonymous label & \graphOnly \\
    E5 Incorrect label & \graphOnly \\
    E6  Inconsistent multi-labeling & \graphOnly \\
    E7 Duplicate entity &  \\
    \quad E7a Exact duplicate entity & S1 Redundancy about an entity \\
    \quad E7b Near duplicate entity & S2 Inconsistency about an entity\\
    E8 Spurious entity & T3 Irrelevant observation\\
    E9 Label-property mismatch & \graphOnly\\
    E10 Inconsistency among property values & T2 Inconsistency among attribute values \\
    E11 Outdated entity & \graphOnly\\
    E12 Outdated label & \graphOnly\\
    \bottomrule
  \end{tabularx}
\end{table}

\paragraph{E1: Missing (connected) entity} A node representing a real-world entity that is expected to occur within the graph scope is absent from the graph instance. 
In the graph excerpt of Figure~\ref{fig:entity-level-graph}, regular patterns in the implicit schema suggest a path connecting an \texttt{Artist} with a \texttt{Track} and \texttt{Work}. However, $n_1$ along with its associated relations $e_1$ and $e_2$ remain missing. Thus, this error implies one or more missing edges in addition to the missing node and can only be detected for connected nodes, as its identification relies on graph context. 
\mapping{This would correspond to a \textit{Missing Tuple} in tabular data. While missing tuples are an important error type even for tabular datasets, they are missing in the taxonomy of Oliveira et al.~\cite{Oliveira2005}, presumably because their detection typically requires external completeness assumptions and is thus difficult to operationalize.}

\paragraph{E2: Missing label} A node or edge lacks a label that is expected for its correct interpretation in the graph. This error type is restricted to single-labeled node entities. 
\mapping{Graph-specific: 
In tabular data tuples cannot exist independently of a table name, thus, such errors only occur in graph data.  
}

\paragraph{E3: Misspelled label} A node has a misspelled label that is required for its correct interpretation in the graph, e.g., \texttt{Atrist} instead of \texttt{Artist}.
\mapping{
Graph-specific: Due to limited schema management, including schema-optional settings, data points within a single type of entity may be unintentionally assigned a misspelled label.
}

\paragraph{E4: Synonymous label} A node has a synonymous label that is required for its correct interpretation in the graph. For example, a customer is labeled as \texttt{Customer} as well as \texttt{Client} across entities carrying the identical property sets.
\mapping{Graph-specific: Analogous to E3, limited schema management increase the risk of such unintended variants. The predefined schema for tabular data prevents such error types by raising a warning. 
}

\paragraph{E5: Incorrect label} A node is assigned a label that does not correctly describe the entity it represents.
\mapping{Graph-specific: Similar to all other label-related error types, tabular data explicitly predefine a schema.
}

\paragraph{E6: Inconsistent multi-labeling} One entity is assigned multiple inconsistent labels, meaning that the same entity receives conflicting or contradictory annotations. The label semantics in $n_2$ imply that the entity \textit{Work} cannot simultaneously be a \textit{Person}.
\mapping{
Graph-specific: Due to the graph-specific context of more than one label this error type has no tabular counterpart.
}

\paragraph{E7: Duplicate entity} The same real-world entity is represented by more than one node. Depending on whether exact duplicates or semantically identical information is contained, two manifestations of this error type arise:
\begin{itemize}
    \item  \textit{E7a: Exact duplicate entity.} Equal or equivalent representation of the same real-world entity. Equality may thereby be caused by a \texttt{copy} operation, while equivalence captures identical semantics, allowing semantic variants in the label, property key or value.
    \mapping{Same as the tabular error type \textit{Redundancy about an entity (S1)}.}
    \item  \textit{E7b: Near duplicate entity.} Inconsistent 
    representation of the same real-world entity. Given an entity is copied with only a subset of the stored key-value pairs or associated edges. 
    \mapping{Same as the tabular error type \textit{Inconsistency about an entity (S2)}.}
\end{itemize}

\paragraph{E8: Spurious entity} An entity that is formally correct but is not relevant in the given context. While in $n_4$ \texttt{James Bond} is a \texttt{Person}, albeit fictional, he is irrelevant in the context of a music graph.
\mapping{Same as the tabular error type \textit{Irrelevant observation (T3)}.}

\paragraph{E9: Label-property mismatch} A node or edge carries a property key, a set of property keys, and their corresponding values that are not compatible with its label. For example, an \texttt{Artist} node carries the property \texttt{duration}.
\mapping{
Graph-specific: Such error types emerge from the tighter coupling of schema and data levels compared to tabular data, paired with the schema optionality. 
}

\paragraph{E10: Inconsistency among property values} There is a violation of an existing dependency between the values of two or more properties. For example, the date value of property \texttt{died} is earlier than that of property \texttt{born}.
\mapping{Same as the tabular error type \textit{Inconsistency among attribute values (T2)}.}

\paragraph{E11: Outdated entity} An entity was not updated accordingly, thus storing an older version.
\mapping{Graph-specific: Analogous to \textit{Outdated property key (P20)}, lazy migration can leave outdated and updated entity representations coexisting in the same graph.}

\paragraph{E12: Outdated label} A label was not updated accordingly. For example, after a schema evolution that renames label \texttt{Band} to \texttt{Group}, some nodes are migrated and labeled \texttt{Group}, while an older node for the same act still carries the outdated label \texttt{Band}.
\mapping{Graph-specific: Analogous to \textit{Missing label (E2)} tabular data tuples cannot exist independently of a table name, thus such errors only occur in graph data.}

\subsection{Relation-level}
\label{subsec:relation-level}
\begin{figure}[htbp]
\centering

\begin{tikzpicture}

\node[pg node] (n1) {
  \textbf{$n_1$: Track}
  \nodepart{second}
  title = 'Back to Black'\\
  duration = 4:01
};

\node[pg node, right = 24.5mm of n1] (n2) {
  \textbf{$n_2$: Work}
  \nodepart{second}
  title = 'Back To Black'\\
  workType = 'song'
};

\node[pg node, below = 18mm of n1] (n3) {
  \textbf{$n_3$: Artist\,\textperiodcentered\,Person}
  \nodepart{second}
  name = 'Amy Winehouse'
};

\node[pg node, below = 18mm of n2] (n4) {
  \textbf{$n_4$: Artist\,\textperiodcentered\,Person}
  \nodepart{second}
  name = 'Mark Ronson'
};

\draw[pg edge]
  (n2) to
  node[pg edge label] {
    \textbf{$e_5$: creates (R3)}\\
    role = 'writer'
  }
  (n4);

\draw[pg edge]
  (n1) to
  node[pg edge label, above, xshift=-0.63mm, yshift=0.57mm] {
    \textbf{$e_4$: }\\
    \textbf{isOriginalRecordingOf}
  }
  (n2);

\draw[pg edge]
  ([xshift = 3.5pt]n3.north) --
  node[pg edge label, right] {
    \textbf{$e_3$: performs}\\
    \textbf{(R2b)}
  }
  ([xshift= 3.5pt]n1.south);

\draw[pg edge]
  ([xshift = -3.5pt]n3.north) --
  node[pg edge label, left] {
    \textbf{$e_2$: performs}\\
    \textbf{(R2b)}\\
    role = 'lead vocals'
  }
  ([xshift = -3.5pt]n1.south);

\draw[pg missing edge]
  (n3) to
  node[pg edge label] {
    \textbf{$e_1$: creates (R1)}\\
    role = 'writer'
  }
  (n2);

\draw[pg edge]
  (n3) to
  node[pg edge label, below] {
    \textbf{$e_6$: likes}\\
    \textbf{(R5)}
  }
  (n4);

\end{tikzpicture}

\caption{Examples of relation-level error types.}
\Description{A figure showing examples of relation-level error types.}
\label{fig:relation-level-graph}

\end{figure}
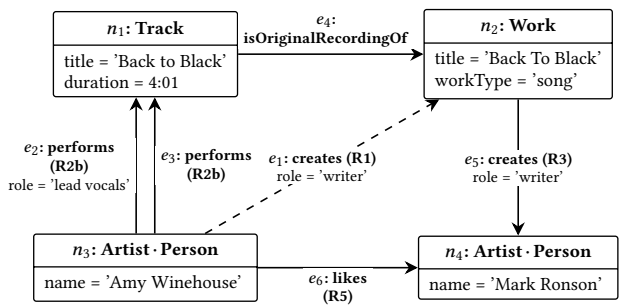

In contrast to tabular data, PGs consist of distinct node and edge elements, whereby their functionality fundamentally differs. 
\textit{Relation-level} errors concern a single relation between two entities and capture inconsistencies arising from relation-specific characteristics. All emerging error types are summarized in Table~\ref{tab:taxonomy-mapping-relation-level} with a subset of them being visualized in Figure~\ref{fig:relation-level-graph}.

\begin{table}[h!]
  \caption{Error taxonomy for relation-level errors in graph data. Entries marked with \graphOnly{} indicate graph-specific error types without a tabular counterpart.}
  \label{tab:taxonomy-mapping-relation-level}

  \small
  \setlength{\tabcolsep}{3pt}
  \renewcommand{\arraystretch}{1.15}

  \begin{tabularx}{\columnwidth}{%
    >{\raggedright\arraybackslash}X
    >{\raggedright\arraybackslash}X
  }
    \toprule
    \textbf{Graph Error Type} & \textbf{Tabular Error Type} \\
    \midrule
    R1 Missing relation & V1 Missing value, \textit{Missing tuple}, T1 Semi-empty tuple \\
    R2 Duplicate relation & \\
    \quad R2a Exact duplicate relation & S1 Redundancy about an entity\\
    \quad R2b Near duplicate relation & S2 Inconsistency about an entity\\
    R3 Incorrect edge direction & \graphOnly \\
    R4 Endpoint type violation & \graphOnly\\
    R5 Spurious relation & \graphOnly \\
    R6 Outdated relation & M2 Outdated Reference\\
    \bottomrule
  \end{tabularx}
\end{table}

\paragraph{R1: Missing relation} A semantic relation between two entities that is expected within the graph scope is absent from the graph instance; illustrated by $e_1$ in Figure~\ref{fig:relation-level-graph}. Such an error directly influences cardinality.
\mapping{This error type cannot be unambiguously assigned to an error type in the tabular taxonomy. The mapping depends on whether tabular data is viewed as a simple table (as in CSV files) or as a relational, normalized database. In a relational, normalized database, relationships are typically represented by foreign keys. In a simple 1-N relationship, this type of error could correspond to a \textit{Missing value (V1)}, more precisely for the column containing the foreign keys. However, in the case of an M-N relationship, a auxiliary table is required in a normalized database. A missing relationship here would be a \textit{Missing tuple} in this auxiliary table. In a CSV file, on the other hand, it is also possible for two entities to appear on the same row. In this case, a missing relationship would correspond to a \textit{Semi-empty tuple (T1)}. In this case, the empty values within the tuple can also occur for foreign-keys, thus directly affecting cardinality.}

\paragraph{R2: Duplicate relation} The same semantic relation of the same source and target. 
This error type is relevant in graph settings in which parallel relation instances occur. Comparable to \textit{Duplicate entity}, this error type is classified based on whether or not duplicates are considered in the strict sense. Thus, two forms can be distinguished:
\begin{itemize}
    \item \textit{R2a: Exact duplicate relation.} Equal or equivalent representation. Equality may thereby be caused by a \texttt{copy} operation, while equivalence captures identical semantics, allowing semantic variants in the label, property key or value.
    \mapping{Same as tabular error type \textit{Redundancy about an entity (S1)}.}
    \item \textit{R2b: Near duplicate relation.} Inconsistent or contradicting representation. A relation that is copied with only a subset of the stored key-value pairs or differs in its source or target entity. 
    \mapping{Same as tabular error type \textit{Inconsistency about an entity (S2)}.}
\end{itemize} 

\paragraph{R3: Incorrect edge direction} A relation is represented with the wrong orientation, so that source and target roles are reversed, e.g., \texttt{creates} edge $e_5$ in Figure~\ref{fig:relation-level-graph}. The participating node types are generally admissible, however, the resulting semantic statement is erroneous.
\mapping{Graph-specific: This error occurs only in graph data because graphs encode relations as directed edges with distinct source and target roles. In relational tabular representations, these roles are fixed by the schema (e.g., separate foreign-key columns in a join table). 
}

\paragraph{R4: Endpoint type violation} A relation connects two entities, but either the source or the target has a type that is inadmissible to the relation type. For example, a \texttt{performs} relation is only admissible between an \texttt{Artist} and a \texttt{Track} node. The participation of two \texttt{Artist} nodes in a \texttt{performs} relation would be inadmissible.
\mapping{
Graph-specific: Graphs explicitly model relations as edges, whereas tabular data relies primarily on key--foreign key relations or auxiliary tables with predefined schemas. Consequently, this error type remains graph-specific. 
}

\paragraph{R5: Spurious relation} A relation is present in the graph although it should not exist. In Figure~\ref{fig:relation-level-graph}, $e_6$ is semantically invalid to describe the relation between the displayed node entities $n_3$ and $n_4$. The relation is beyond the scope of a music graph.
\mapping{
Graph-specific: The modeling freedom in graphs allow such semantic heterogeneity that requires external context of the domain to be identified. It cannot occur in tabular data because this would require the intentional creation of either a key--foreign key relation or an auxiliary table.
}

\paragraph{R6: Outdated relation} A relation was not updated accordingly.
\mapping{Same as tabular error type \textit{Outdated reference (M2)}.}

\subsection{Pattern-level}
\label{subsec:pattern-level}
While all previous levels focus on error types centered around single elements, more complex structures require considering context, topology and connectivity (see Section~\ref{sec:dq-challenges}). In graph data, single elements may not reveal the full error in isolation, while their joint configuration violates an expected structure or semantics. Such configurations form meaningful patterns and thus necessitate consideration beyond single graph elements. 
To capture errors within such patterns, we define complex error types for them. Complex error types represent semantic sets of atomic errors whose graph-specific relevance emerges from their joint configuration. Nevertheless, each complex error can be decomposed into their underlying atomic error types.
\textit{Pattern-level} errors begin where the violated condition can only be stated over a multi-element configuration, such as a participation structure, a path, a cycle, or a hierarchy. Table~\ref{tab:taxonomy-mapping-pattern-level} provides an overview of all pattern-level error types.

\begin{table}[h!]
  \caption{Error taxonomy for pattern-level in graph data. Entries marked with \graphOnly{} indicate graph-specific error types without a tabular counterpart.}
  \label{tab:taxonomy-mapping-pattern-level}

  \small
  \setlength{\tabcolsep}{3pt}
  \renewcommand{\arraystretch}{1.15}

  \begin{tabularx}{\columnwidth}{%
    >{\raggedright\arraybackslash}X
    >{\raggedright\arraybackslash}X
  }
    \toprule
    \textbf{Graph Error Type} & \textbf{Tabular Error Type} \\
    \midrule
    Pa1 Participation or cardinality error & \graphOnly \\
    Pa2 Structural configuration error & \textit{Value depends structure FD }\\
    \quad Pa2a Acyclicity violation & \graphOnly\\
    \quad Pa2b Root or parent error & \graphOnly \\
    \quad Pa2c Symmetry or inverse violation & \graphOnly \\
    Pa3 Duplicate pattern &  \\
    \quad Pa3a Exact duplicate pattern & \graphOnly\\
    \quad Pa3b Near duplicate pattern & \graphOnly\\
    Pa4 Inconsistent pattern & \graphOnly \\
    Pa5 Conflated neighborhood & \graphOnly \\
    \bottomrule
  \end{tabularx}
\end{table}

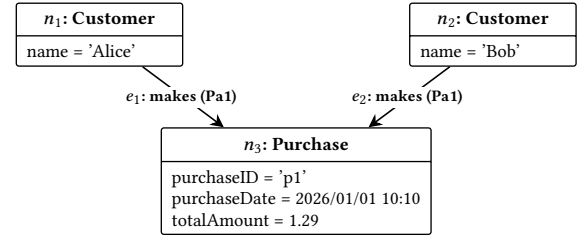
\begin{figure}[htbp]
\centering

\begin{tikzpicture}

\node[pg node] (n3) {
  \textbf{$n_3$: Purchase}
  \nodepart{second}
  purchaseID = 'p1'\\
  purchaseDate = 2026/01/01 10:10\\
  totalAmount = 1.29
};

\node[pg node, above left= 8mm and -3mm of n3] (n1) {
  \textbf{$n_1$: Customer}
  \nodepart{second}
  name = 'Alice'
};

\node[pg node, above right=8mm and -3mm of n3] (n2) {
  \textbf{$n_2$: Customer}
  \nodepart{second}
  name = 'Bob'
};

\draw[pg edge]
  (n1) to
  node[pg edge label] {
    \textbf{$e_1$: makes (Pa1)}
  }
  (n3);

\draw[pg edge]
  (n2) to
  node[pg edge label] {
    \textbf{$e_2$: makes (Pa1)}
  }
  (n3);

\end{tikzpicture}

\caption{Cardinality error (Pa1).}
\Description{A figure showing an example of an cardinality error.}
\label{fig:cardinality-graph}

\end{figure}

\paragraph{Pa1: Participation or cardinality error} A node or set of nodes participates in too few or too many relation instances of a given type, relative to an applicable expectation. The error is therefore not exhausted by the presence or absence of one particular edge, but concerns the participation structure as a whole. For example, a \texttt{Purchase} is supposed to have a 1-1 relationship with a \texttt{Customer}, however, in Figure~\ref{fig:cardinality-graph} two different customers, namely $n_1$ and $n_2$ are associated with the same \texttt{Purchase} node through two ingoing \texttt{makes} labeled edges, resulting in a cardinality error. 
\mapping{Graph-specific: In contrast to tabular data, graphs do not explicitly define cardinality within the predefined schema. Since cardinality constraints are implicitly modeled in graphs and directly affect patterns including their semantics, cardinality violations are located at the pattern-level.}

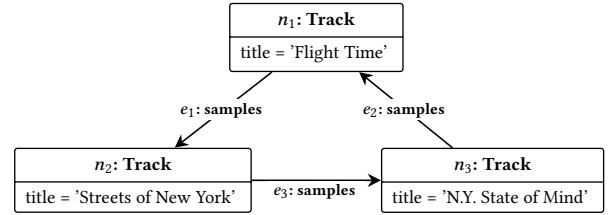
\begin{figure}[htbp]
\centering

\begin{tikzpicture}

\node[pg node] (n1) {
  \textbf{$n_1$: Track}
  \nodepart{second}
  title = 'Flight Time'
};

\node[pg node, below left=10mm and -3mm of n1] (n2) {
  \textbf{$n_2$: Track}
  \nodepart{second}
  title = 'Streets of New York'
};

\node[pg node, below right=10mm and -3mm of n1] (n3) {
  \textbf{$n_3$: Track}
  \nodepart{second}
  title = 'N.Y. State of Mind'
};

\draw[pg edge]
  (n3) to
  node[pg edge label] {
    \textbf{$e_2$: samples}
  }
  (n1);

\draw[pg edge]
  (n2) to
  node[pg edge label, below] {
    \textbf{$e_3$: samples}
  }
  (n3);

\draw[pg edge]
  (n1) to
  node[pg edge label] {
    \textbf{$e_1$: samples}
  }
  (n2);

\end{tikzpicture}

\caption{Acyclicity violation (Pa2a).}
\Description{A figure showing an example of an acyclicity violation.}
\label{fig:acyclicity-graph}

\end{figure}

\paragraph{Pa2a: Acyclicity violation} A subgraph contains a cycle although the relevant relation type or modeling convention requires acyclicity. For example, (in Figure~\ref{fig:acyclicity-graph}) several music tracks form a cycle through the \texttt{samples} relation type, although necessarily one of these tracks has to be the oldest thus being unable to sample a track that has been created later.
\mapping{Graph-specific: Even though such errors result from the inherit structure of graphs, \textit{Circularity among tuples in a self-relationship (M5)} represent a similar, yet different error type in tabular data. Such an tabular error is defined as circularity between two distinct tuples. However, in graph data circularity can occur either within data points of the same or multiple different labels, thus increasing the complexity of this error type compared to tabular data.}

\begin{figure}[htbp]
\centering

\begin{tikzpicture}

\node[pg node] (n1) {
  \textbf{$n_1$: Release}
  \nodepart{second}
  title = 'Casino Royale'\\
  releaseType = 'audiobook'
};

\node[pg node, above=8mm of n1] (n2) {
  \textbf{$n_2$: Genre}
  \nodepart{second}
  name = 'Spy thriller'
};

\node[pg node, above left=8mm and -2mm of n2] (n3) {
  \textbf{$n_3$: Genre}
  \nodepart{second}
  name = 'Thriller'
};

\node[pg node, above right=8mm and -2mm of n2] (n4) {
  \textbf{$n_4$: Genre}
  \nodepart{second}
  name = 'Fiction'
};

\draw[pg edge]
  (n1) to
  node[pg edge label,below] {
    \textbf{$e_1$: hasGenre}
  }
  (n2);

\draw[pg edge]
  (n2) to
  node[pg edge label] {
    \textbf{$e_2$: isSubgenreOf}
  }
  (n3);
  
\draw[pg edge]
  (n2) to
  node[pg edge label] {
    \textbf{$e_3$: isSubgenreOf}
  }
  (n4);

\end{tikzpicture}

\caption{Root/Parent violation (Pa2b).}
\Description{A figure showing an example of a root violation.}
\label{fig:root-parent-graph}

\end{figure}
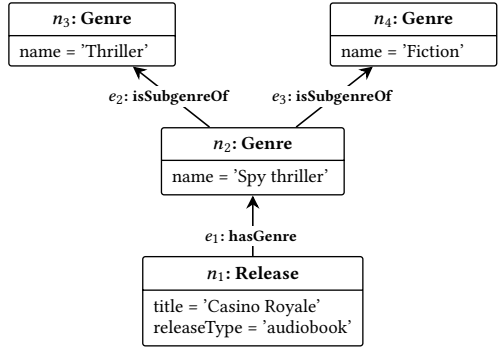

\paragraph{Pa2b: Root or parent error} A subgraph intended to form a rooted hierarchy fails a requirement, because a node lacks a required parent or has several parents although a single parent is intended. In Figure~\ref{fig:root-parent-graph} the value \texttt{audiobook} of the \texttt{releaseType} property determines a specific structure. More precisely, it requires a connection to a \texttt{Genre} node with the key-value pair \textit{name = 'Fiction'} or \textit{name = 'Non-Fiction'} at its root. 
\mapping{Graph-specific: This pattern resembles a specific manifestation of a functional dependency within graph data. Similar to JSON data~\cite{DBLP:conf/edbt/KlessingerFGKSS23}, this represents an instance-specific pattern which is conditioned on a property key's value.} 

\begin{figure}[htbp]
\centering

\begin{tikzpicture}

\node[pg node] (n1) {
  \textbf{$n_1$: Artist\,\textperiodcentered\,Person}
  \nodepart{second}
  name = 'Lou Reed'
};


\node[pg node, below=10mm of n1] (n3) {
  \textbf{$n_2$: Artist\,\textperiodcentered\,Group}
  \nodepart{second}
  name = 'The Velvet\\ \phantom{name = '}Underground'
};

\node[pg node, right= 19.5mm of n3] (n4) {
  \textbf{$n_3$: Artist\,\textperiodcentered\,Person}
  \nodepart{second}
  name = 'Nico'
};



\draw[pg edge]
  ([xshift=3.5pt]n3.north) --
  node[pg edge label, right] {
    \textbf{$e_1$: isMemberOf}
  }
  ([xshift=3.5pt]n1.south);

\draw[pg edge]
  ([xshift=-3.5pt]n1.south) --
  node[pg edge label, left] {
    \textbf{$e_2$: isMemberOf}
  }
  ([xshift=-3.5pt]n3.north);

\draw[pg missing edge]
  ([yshift=-3.5pt]n4.west) --
  node[pg edge label, below] {
    \textbf{$e_3$: collaborates}
  }
  ([yshift=-3.5pt]n3.east);

\draw[pg edge]
  ([yshift=3.5pt]n3.east) --
  node[pg edge label, above] {
    \textbf{$e_4$: collaborates}
  }
  ([yshift=3.5pt]n4.west);

\end{tikzpicture}

\caption{Symmetry and Inverse violation (Pa2c).}
\Description{A figure showing an example of a symmetry and inverse violation.}
\label{fig:symmetry-inverse-graph}

\end{figure}
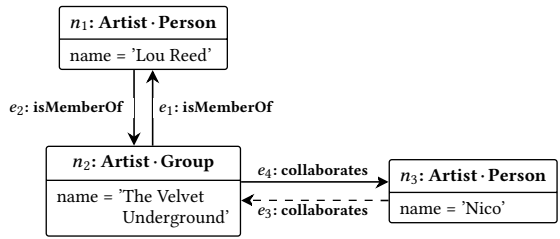

\paragraph{Pa2c: Symmetry or inverse violation} A relation fails to satisfy an expected dependency between relation instances, because a required anti-symmetric counterpart is missing or an inadmissible reverse edge exists although symmetry is required. In Figure~\ref{fig:symmetry-inverse-graph}, a \texttt{collaborates} relation is inherently symmetric, while \texttt{isMemberOf} and \texttt{hasMember} form a pair of inverse relation types. 
\mapping{Graph-specific: In graph data, edges are directed, thus, they may require corresponding edges to be modeled in case of symmetry, requiring constraint creation to ensure a mandatory modeling if needed. Hence, this error types refers to a non-defined existence constraint, however schema optionality does enforce the declaration of explicit existence constraints.}

\paragraph{Pa3: Duplicate pattern} 

Multiple realizations of a graph pattern encode the same semantic statement about the same focal group of entities within the same validity context.
The entry is reserved for redundancies that cannot be adequately reduced to a mere duplicate node or duplicate edge, but instead concerns an entire structural subgraph. It may consist of multiple entities and their connecting relationships, including associated labels or key-value pairs.
Two forms can be distinguished:
\begin{itemize}
    \item \textit{Pa3a: Exact duplicate pattern.} Equal or equivalent representation of a complete graph or subgraph while preserving its structure and relevant semantic characteristics. 
    Equality may result from a \texttt{copy} operation, while equivalence captures identical semantics, allowing semantic variants in the labels, property keys, or values.
    \mapping{Graph-specific: Intended redundancies can be used as an optimization strategy or as backup when schema changes are required.}
    \item \textit{Pa3b: Near duplicate pattern.} A graph pattern only partially corresponds to the original realization. This may result from copying a pattern under selection, projection or filtering conditions, such that only a subset of entities, relations, labels or key-value pairs is retained. 
    \mapping{Graph-specific: Such heterogeneity can emerges when parts of the graph are duplicated for different purposes, while retaining only a subset of the original nodes, edges and key-value pairs.}
\end{itemize} 

\begin{figure}[htbp]
\centering

\begin{tikzpicture}

\node[pg node] (n1) {
  \textbf{$n_1$: Track}
  \nodepart{second}
  title = 'Skyfall'\\
  duration = 5:18
};

\node[pg node, above left=8mm and 3.5mm of n1] (n2) {
  \textbf{$n_2$: Work}
  \nodepart{second}
  workID = 'w1'\\
  title = 'Skyfall'\\
  workType = 'song'
};

\node[pg node, above right=8mm and 3.5mm of n1] (n3) {
  \textbf{$n_3$: Work}
  \nodepart{second}
  workID = 'w2'\\
  title = 'Skyfall'\\
  workType = 'song'
};

\node[pg node, below left=8mm and 2mm of n1] (n4) {
  \textbf{$n_4$: Artist}
  \nodepart{second}
  name = 'Travis Scott'
};

\node[pg node, below right=8mm and 3.5mm of n1] (n5) {
  \textbf{$n_5$: Artist}
  \nodepart{second}
  name = 'Adele'
};

\draw[pg edge]
  (n4) to
  node[pg edge label] {
    \textbf{$e_5$: performs}\\
    role = 'vocals'
  }
  (n1);

\draw[pg edge]
  (n5) to
  node[pg edge label, right] {
    \textbf{$e_4$: creates}\\
    role = 'writer'
  }
  (n3);

\draw[pg edge]
  (n4) to
  node[pg edge label, left] {
    \textbf{$e_3$: creates}\\
    role = 'writer'
  }
  (n2);

\draw[pg edge]
  (n1) to
  node[pg edge label] {
    \textbf{$e_2$:  isCoverOf}
  }
  (n3);

\draw[pg edge]
  (n1) to
  node[pg edge label] {
    \textbf{$e_1$: isOriginalRecordingOf}
  }
  (n2);

\end{tikzpicture}

\caption{Inconsistent pattern (Pa4).}
\Description{A figure showing an example of an inconsistent pattern.}
\label{fig:pa-inconsistent}

\end{figure}
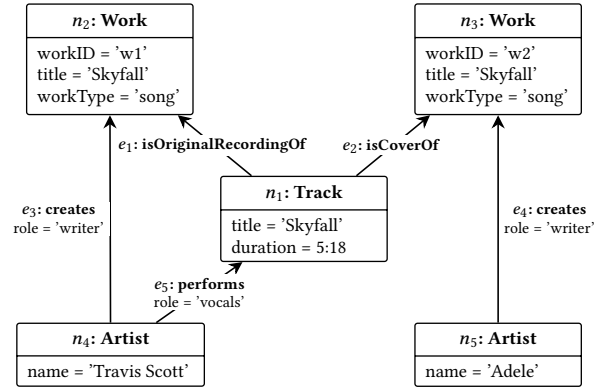

\paragraph{Pa4: Inconsistent pattern} Two or more subgraphs concerning the same focal entity or set of entities jointly express mutually exclusive or otherwise inconsistent statements. In Figure~\ref{fig:pa-inconsistent}, a music \texttt{Track} is an original recording of a \texttt{Work} of music embedded in its own neighborhood, yet it is connected to another writer's \texttt{Work} of the same name by an \texttt{isCoverOf} relation $e_2$, signifying that it is a cover version of it.
\mapping{Graph-specific: The highly interconnected nature of graph data paired with the schema optionality allows contradictory or exclusive statements to be modeled.}

\paragraph{Pa5: Conflated neighborhood} One node holds information about more than one real-world entity, resulting in connectivity changes. This manifests as both connections that should not be present in the graph and as missing relations. In Figure~\ref{fig:pa-conflated-graph} information on two different releases is stored in $n_1$, thus conflating two distinct neighborhoods and losing information on the \texttt{releaseYear} of the Adele single 'Hello' and the \texttt{trackCount} of the Lionel Richie single 'Hello'.
Besides information loss, the unintended merging of two distinct nodes and subsequently their neighborhoods results in a higher connectivity of the affected node, thus indicating higher importance within the graph. Furthermore, it is transformed into a bridging node between parts of a graph, thereby altering shortest paths~\cite{DBLP:journals/access/LiuPW19}. For instance, \texttt{Lionel Richie} is now reachable within two hops from the \texttt{XL} record label, although he was never signed to that label. Moreover, missingness emerges from the unintended merging which otherwise would be modeled within the graph; indicated by the dashed graph elements. 
\mapping{Graph-specific: The modeling flexibility of graphs coupled with the limited support of schema management increases the risk of unintended outcomes when merging or working in collaborative environments. In case of a conflated neighborhood connectivity is affected since parts are missing, false relations emerge and can cause unintended hubs to emerge.}

\definecolor{mutedRed}{RGB}{181, 101, 91}

\begin{figure}[htbp]
\centering

\begin{tikzpicture}

\node[pg node] (n1) {
  \textbf{$n_1$: Release}
  \nodepart{second}
  title = 'Hello'\\
  releaseType = 'single'\\
  trackCount = 1\\
  releaseYear = 1984
};

\node[pg node, above=6mm and -1mm of n1] (n2) {
  \textbf{$n_2$: Artist}
  \nodepart{second}
  name = 'Adele'
};

\node[pg node, below=6mm and -1mm of n1] (n3) {
  \textbf{$n_3$: Label}
  \nodepart{second}
  name = 'XL'
};

\node[pg node, above right=6mm and 7mm of n1] (n4) {
  \textbf{$n_4$: Artist}
  \nodepart{second}
  name = 'Lionel Richie'
};

\node[pg node, below right=6mm and 7mm of n1] (n5) {
  \textbf{$n_5$: Label}
  \nodepart{second}
  name = 'Motown'
};

\node[pg missing node, right=30mm of n1] (n6) {
  \textbf{$n_6$: Release}
  \nodepart{second}
  title = 'Hello'\\
  releaseType = 'single'\\
  \textit{trackCount = 2}\\
  releaseYear = 1984
};

\draw[pg edge, mutedRed]
  (n4) to
  node[pg edge label] {
    \textbf{$e_1$: isCreditedOn}
  }
  (n1);

\draw[pg edge]
  (n2) to
  node[pg edge label] {
    \textbf{$e_3$: isCreditedOn}
  }
  (n1);
  
\draw[pg edge]
  (n3) to
  node[pg edge label] {
    \textbf{$e_2$: issues}
  }
  (n1);

\draw[pg edge, mutedRed]
  (n5) to
  node[pg edge label] {
    \textbf{$e_4$: issues}
  }
  (n1);

\draw[pg edge, dashed]
  (n4) to
  node[pg edge label] {
    \textbf{$e_5$: isCreditedOn}
  }
  (n6);

\draw[pg edge, dashed]
  (n5) to
  node[pg edge label] {
    \textbf{$e_6$: issues}
  }
  (n6);

\end{tikzpicture}

\caption{Conflated neighborhood (Pa5). Dashed graph elements represent missingness resulting from this error, while color-coded graph elements should not be present in the graph.}
\Description{A figure showing an example of an conflated neighborhood.}
\label{fig:pa-conflated-graph}

\end{figure}
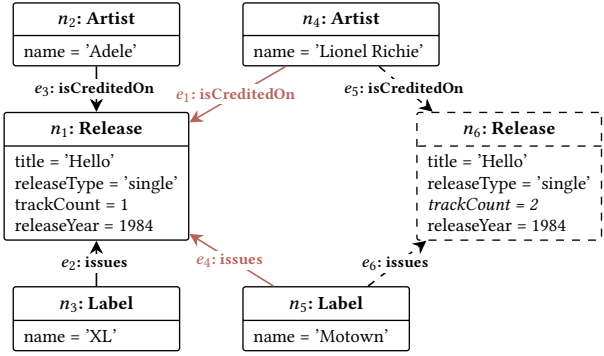

\subsection{Graph-level}
\label{subsec:graph-level}
Errors that can only be identified by considering the entire graph, including its elements, structures, and semantics, are located at the \textit{Graph-Level} (Section~\ref{subsec:graph-level}). Even though complex pattern- and graph level error types can be decomposed into atomic errors, they represent semantic error sets, describing a specific combination of those. An overview of the graph-level error types is given in Table~\ref{tab:taxonomy-mapping}.

\begin{table}[ht]
  \caption{Error taxonomy for graph-level errors in graph data.}
  \label{tab:taxonomy-mapping}

  \small
  \setlength{\tabcolsep}{3pt}
  \renewcommand{\arraystretch}{1.15}

  \begin{tabularx}{\columnwidth}{%
    >{\raggedright\arraybackslash}X
    >{\raggedright\arraybackslash}X
  }
    \toprule
    \textbf{Graph Error Type} & \textbf{Tabular Error Type} \\
    \midrule
    G1 Bias & S3 Bias \\
    G2 Noise & S4 Noise \\
    G3 Syntax inconsistency & M3 Syntax inconsistency \\
    \bottomrule
  \end{tabularx}
\end{table}

\paragraph{G1: Bias} Systematic distortion in the data that compromises its representativeness.
\mapping{Same as the tabular error type \textit{Bias (S3)}.}

\paragraph{G2: Noise} The graph-wide ratio of error-free to erroneous data exceeds a tolerated threshold. Unlike bias, this refers to non-systematic data collection errors.
\mapping{Same as the tabular error type \textit{Noise (S4)}.}

\paragraph{G3: Syntax inconsistency} Depending on the entity, there are different representations for properties that have the same type.
\mapping{Same as the tabular error type \textit{Syntax inconsistency (M3)}.}

\subsection{Tabular-specific Error Types}
Building on the comprehensive taxonomy for graph data and the corresponding mapping to established tabular error notions, our analysis confirms that a subset of data quality issues is inherently graph-specific, driven by model characteristics such as schema optionality and semantics induced by connectivity. At the same time, the mapping process is informative in the opposite direction: it reveals that some quality concepts presuppose table-internal constructs (e.g., explicit keys and foreign-key constraints) and therefore do not have a direct counterpart in PGs. The resulting tabular-specific error types include:

\paragraph{Referential integrity violation (M1)} This type of error would most closely correspond to a \textit{dangling edge} -- that is, an edge in a graph whose source or target node is missing or no longer exists. However, this is not possible to model in graphs. In tables, on the other hand, it is possible, for example, to reference a primary key that does not exist. Accordingly, this type of error can only occur in tabular data.

\paragraph{Inconsistency among related attribute values (M4)} This error type is defined through inter-relational dependencies and aggregation semantics (e.g., comparisons between values and sums over foreign-key related tuples). Such semantics are not natively encoded or enforced in PGs and depend on application-specific subgraph patterns and traversal scopes. While one could model this as a pattern-level constraint over a particular graph neighborhood, we exclude it from our technology-independent taxonomy because its formulation is inherently model- and query-dependent.

\paragraph{Circularity among tuples in a self-relationship (M5)} As described, this error is conceptually related to \textit{Acyclicity violation (Pa2a)}, but it is defined more narrowly in the tabular setting: it refers to cyclic references induced by a self-relationship between distinct tuples. In PGs, analogous cycles may span heterogeneous node and edge types and can involve multiple labels and relationship kinds, which makes the corresponding violation less localized.

\paragraph{Domain constraint violation (M6)} Oliveira et al.~\cite{Oliveira2005} originally define \textit{domain constraint violation} as an error type on each structural level. In our definition~\cite{GouDa}, we deliberately excluded this category because in practice, a domain constraint violation is typically a domain-specific instantiation of other error types (e.g., \textit{Syntax violation (V2)}). Consequently, we do not include \textit{domain constraint violation} as a separate category in the graph error taxonomy. For PGs, such domain rules are usually not natively encoded or enforced, since schema optionality offers a greater modeling flexibility without predefining a schema. Instead, they are formulated and checked as application-specific pattern constraints whose exact specification depends on modeling choices (labels, relation types, and neighborhoods). This prevents a model-independent mapping to a standalone graph error category.

\section{Discussion}
\label{sec:discussion}
The presented error taxonomy captures graph error types across five structural levels with increasing complexity and ensures general applicability to any PG. During the taxonomy design all data quality issues, shown in Section~\ref{sec:dq-challenges}, were considered to ensure a comprehensive analysis of how errors manifest in the underlying graph data, including the effects of connectivity, context and schema optionality. Section~\ref{sec:errors} outlines in detail how tabular error types increase in complexity within the graph model, resulting in atomic (property-, entity- and relation-levels) and complex error types (pattern- and graph-levels). Although our contribution focuses on the taxonomy, this is an essential step, as it provides the foundation for subsequent error detection and repair. We thus revisit key characteristics and challenges of graph data and outline how they shape the design of detection and repairing approaches. 

\paragraph{Schema Optionality} Since tabular data heavily relies on a predefined schema, mandatory graph-specific extensions are directly linked to the flexible schema and the pattern-focus emerging from interconnection. This structure 
also affects how the data itself are modeled. In graph data, representing properties as key-value pairs attached to individual graph elements, together with the schema-relevant information such as labels or edges, fundamentally differs from the structure of tabular data. As a result, tabular error types are heavily centered around the data level, while implicit or incompletely defined graph schemas can cause cross-layer error types concerning both the schema and data levels simultaneously.

\paragraph{Cross-Level Heterogeneity} Error detection relies on deriving regularities from the implicit schema to identify irregularities. The modeling flexibility in graph data allows variations to occur across all discussed levels. Besides spanning all levels, heterogeneity can manifest in multiple forms, such as syntactic variations, semantical differences, or structural irregularities. Especially, when occurring in combination, their detection is complicated. Nevertheless, heterogeneity does not necessarily represent an error rather than an intended variation, thus requiring validation through external context.

\paragraph{Structural Semantics} Additionally, error types manifest in other forms, since the connectivity of graph data encodes rich semantics that can span large subgraphs or even the entire graph. 
Instead, semantics in tabular data typically are more localized within attributes and tuples.
However, structural semantics additionally emerge from the graph topology itself, while schema-relevant information such as labels and property keys is attached directly to individual graph elements. This illustrates how, in PGs, schema information is more closely intertwined with the data itself. Therefore, PG error types cannot always be characterized by the schema and data dimensions in isolation.

\paragraph{Data Dependencies} This difference between tabular data and PGs also becomes apparent when considering dependencies between data values, since they are typically more explicitly represented in tabular data. Paired with the modeling flexibility of graphs, the same information can be stored with different degrees of normalization within a single PG instance. Identifying dependencies between such representations is therefore important for determining whether redundant or related values remain consistent. This depicts the need to include dependency detection into future error detection tools to detect errors arising from different degrees of normalization, such as a \textit{Root or Parent Error}. Such errors may subsequently require handling through schema evolution operations, such as \textit{split} operations.

\paragraph{Schema Evolution \& Versioning.} Besides that, the schema optionality of NoSQL databases makes them well suited to lazy data migration approaches where only a subset of affected data points is updated~\cite{DBLP:journals/corr/ScherzingerKS13}. This can result in entire graph structures or subsets of information being stored in multiple versions simultaneously \cite{DBLP:journals/rcs/BoukettayaNG20}. 
The coexistence of outdated and updated versions within the same graph can lead to temporal irregularities. Errors resulting from lazy migration require external context to be reliably classified (e.g., \textit{Outdated property key (P20)} or \textit{Outdated entity (E11)}).

\paragraph{Technology-Independence and Graph Model Scope} Our taxonomy is based on a technology-independent perspective, consistent with standardization efforts, which likewise define PG concepts independently of a specific underlying database system to support applicability across implementations. Although PGs and RDF graphs both represent interconnected data, they rely on different modeling techniques. RDFs graphs store information in subject, predicate, object triples, whereas PGs distinguish graph elements, namely nodes, edges, and properties. This directly influences the manifestation of errors and requires further consideration, since errors may not necessarily be directly mappable between the two graph models. 
Moreover, the formal semantics of RDF affect how missingness, irregularities, or conflicts are interpreted in the context of error detection. Hence, our PG-specific taxonomy provides a foundation that can be extended and refined in future work to account for error types in other graph models.
\par\smallskip

One of the key challenges encountered during the mapping between tabular and graph errors is the absence of tabular counterparts for a subset of graph errors, which consequently constitute graph-specific error types. Besides the blurred distinction between the schema and data levels, the identified challenges, presented in Section~\ref{sec:dq-challenges}, show that graph errors cannot always be characterized using concepts transferred directly from tabular data. This results from the inherent modeling of graph data, where schema-related information and data values are jointly represented through graph elements, labels and key-value pairs and surroundings contain rich semantics. In addition, PGs require a graph-specific organization of error types across structural levels.  

Overall, our findings show that a non-trivial subset of data quality issues is inherently graph-specific, primarily due to schema optionality and the rich semantics induced by connectivity. These characteristics not only motivate the graph-specific structural organization of our taxonomy, but also imply that detection and repair often require context beyond single values, ranging from neighborhoods to subgraphs.

\section{Conclusion and Future Work}
\label{sec:conc}
In this paper, we studied data quality in the context of graph data and developed a comprehensive error taxonomy. We further derived a mapping between errors in graph data and corresponding error notions in tabular data. This mapping connects established concepts from tabular data quality to graph-specific characteristics. It clarifies which error concepts transfer across data models and where graph data introduce distinct challenges. To the best of our knowledge, this is the first study to provide such a detailed taxonomy and mapping for errors in graph data.

While this paper focused on the error taxonomy, a natural next step is to develop detection and repair techniques tailored to graph-specific challenges such as schema optionality, structural semantics, and heterogeneous normalization within a single PG. In particular, future work should investigate (i) constraint- and dependency-based detection methods, and (ii) repair strategies that explicitly account for schema evolution and versioning (e.g., coexisting outdated and updated representations).

\begin{acks}
The work of Dominique Hausler has been funded by Deutsche Forschungsgemeinschaft (German Research Foundation) -- 552691270.
\end{acks}

\section*{Declaration on Generative AI}
ChatGPT (GPT-5.6 Sol) was used for rephrasing and as a thesaurus.

\section*{Artifacts}
A comprehensive graph that takes all error types of the PG error taxonomy into account is available on Zenodo: \url{https://doi.org/10.5281/zenodo.23207484}. This includes highlighting and encoding of all error types in alignment with this paper.

\balance
\bibliographystyle{ACM-Reference-Format}
\bibliography{bibliography}

\end{document}